\documentclass[psf,proceedingpaper,accept,pdftex,moreauthors]{Definitions/mdpi} 
\usepackage{caption}
\usepackage[labelformat=simple]{subcaption}

\DeclareCaptionLabelFormat{subcaptionlabel}{\normalfont(\textbf{#2}\normalfont)}
\usepackage{todonotes}
\usepackage{verbatim, bm}
\usepackage{xcolor}
\firstpage{1} 
\pubvolume{9}
\issuenum{1}
\articlenumber{10}
\pubyear{2023}
\copyrightyear{2023}
\externaleditor{Academic Editors: Udo von Toussaint and Roland Preuss}
\datepublished{29 November 2023} 
\hreflink{https://doi.org/10.3390/\linebreak psf2023009010} 
\doinum{10.3390/psf2023009010}

\Title{Learned Harmonic Mean Estimation of the Marginal Likelihood with Normalizing Flows $^{\dagger}$}

\TitleCitation{Learned Harmonic Mean Estimation of the Marginal Likelihood with Normalizing Flows}

\Author{Alicja Polanska $^{1,}$*, Matthew A. Price $^{1}$, Alessio Spurio Mancini $^{1}$ and Jason D. McEwen $^{1,2}$}

\AuthorNames{Alicja Polanska, Matthew A. Price, Alessio Spurio Mancini and Jason D. McEwen}

\AuthorCitation{Polanska, A.; Price, M.A.; Spurio Mancini, A.; McEwen, J.D.}

\address{%
$^{1}$ \quad Mullard Space Science Laboratory, University College London, Dorking RH5 6NT, UK; m.price.17@ucl.ac.uk~(M.A.P.); a.spuriomancini@ucl.ac.uk (A.S.M.); jason.mcewen@ucl.ac.uk (J.D.M.)\\
$^{2}$ \quad Alan Turing Institute, London NW1 2DB, UK}

\corres{Correspondence: alicja.polanska.22@ucl.ac.uk}

 \firstnote{Presented at the 42nd International Workshop on Bayesian Inference and Maximum Entropy Methods in Science and Engineering, Garching, Germany, 3–7 July 2023.} 

\abstract{Computing the marginal likelihood  (also called the Bayesian model evidence) is an important task in Bayesian model selection, providing a principled quantitative way to compare models.  The learned harmonic mean estimator solves the exploding variance problem of the original harmonic mean estimation of the marginal likelihood. The learned harmonic mean estimator learns an importance sampling target distribution that approximates the optimal distribution. While the approximation need not be highly accurate, it is critical that the probability mass of the learned distribution is contained within the posterior in order to avoid the exploding variance problem.  In previous work, a bespoke optimization problem is introduced when training models in order to ensure this property is satisfied.  In the current article, we introduce the use of normalizing flows to represent the importance sampling target distribution.  A flow-based model is trained on samples from the posterior by maximum likelihood estimation. Then, the probability density of the flow is concentrated by lowering the variance of the base distribution, i.e., by lowering its ``temperature'', ensuring that its probability mass is contained within the posterior.  This approach avoids the need for a bespoke optimization problem and careful fine tuning of parameters, resulting in a more robust method.  Moreover, the use of normalizing flows has the potential to scale to high dimensional settings.  We present preliminary experiments demonstrating the effectiveness of the use of flows for the learned harmonic mean estimator.  The harmonic code implementing the learned harmonic mean, which is publicly available, has been updated to now support normalizing flows.
}


\begin{document}
\newcommand{\prob}{\ensuremath{{p}}}
\newcommand{\evidence}{\ensuremath{{z}}}
\newcommand{\evidenceinv}{\ensuremath{\rho}}
\newcommand{\data}{\ensuremath{{y}}}
\newcommand{\datum}{\ensuremath{y}}
\renewcommand{\exp}{\ensuremath{\text{exp}}}
\newcommand{\given}{\ensuremath{{\,\vert\,}}}

\section{Introduction}
Model selection is the task of evaluating which of the statistical models under consideration best describes observed data. In~the Bayesian formalism, this involves computing the marginal likelihood (also called the Bayesian model evidence), which gives a way to a quantitative comparison of the suitability of models for a given problem. Model selection is relevant in a range of fields such as astronomy, biostatistics, economics, medical research, and many more. However, in~practice, Bayesian model selection is often difficult, as computing the marginal likelihood requires evaluating a high-dimensional integral, which can be a challenging task. A~number of methods to compute the marginal likelihood have been proposed (for reviews see~\cite{clyde2007current,friel2012estimating}), such as nested sampling~\cite{skilling2006nested,ashton2022nested}.

The learned harmonic mean estimator was proposed recently by some of the authors of the current article as an effective technique with which to compute the marginal likelihood~\cite{mcewen2021machine}.  The~estimator requires only samples from the posterior and so is agnostic to the method used to generate samples, in~contrast to nested sampling.  Thus, the~learned harmonic mean can be easily coupled with various MCMC sampling techniques or variational inference approaches.  This property also allows the estimator to be adapted to address Bayesian model selection for simulation-based inference (SBI) \cite{mancini2022bayesian},
where an explicit likelihood is unavailable or infeasible. The~learned harmonic mean estimator is implemented in the harmonic Python code (\url{https://github.com/astro-informatics/harmonic}, accessed on 30 June 2023).

In this work, we introduce the use of normalizing flows~\cite{papamakarios2021normalizing} for the learned harmonic mean estimator, which addresses some limitations of the models used previously.   The~use of normalizing flows eliminates the need for bespoke training, resulting in a more robust and scalable approach, which is now implemented in the harmonic code.  We first review the learned harmonic mean estimator, before~describing how normalizing flows may be used for the estimator and their main advantages in this context.  We then present a number of experiments that demonstrate the effectiveness of the use of flows in the learned harmonic mean~estimator.

\section{The Harmonic Mean~Estimator}
Bayesian model selection requires the computation of the marginal likelihood given~by
\begin{equation}
	z =
	\prob(\data \given M)
	= \int \text{d} \theta \:
	\prob(\data \given \theta, M) \: \prob(\theta \given M)
	= \int \: \text{d} \theta \mathcal{L}(\theta) \pi(\theta) ,
\end{equation}
where $\data$ denotes observed data, $\theta$ the parameters of interest, and~$M$ the model under consideration.  We adopt the shorthand notation for the likelihood of $\mathcal{L}(\theta) = \prob(y \given \theta, M)$ and the prior of $\pi(\theta) = \prob(\theta \given M)$.

The harmonic mean estimator was first proposed by~\cite{newton1994approximate}, who showed that the marginal likelihood $z$ can be estimated from the harmonic mean of the likelihood, given posterior samples.  This follows by considering the expectation of the reciprocal of the likelihood with respect to the posterior distribution, leading to the following estimator of the reciprocal of the marginal likelihood $\rho = 1/z$:
\vspace{-3pt}
\begin{equation}
	\hat{\rho} = \frac{1}{N} \sum_{i=1}^{N} \frac{1}{\mathcal{L}(\theta_i)} ,
	\quad
	\theta_i \sim \prob(\theta \given \data) ,
\end{equation}
where $N$ specifies the number of samples $\theta_i$ drawn from the posterior $\prob(\theta \given \data)$. The~marginal likelihood can then be estimated from its reciprocal straightforwardly~\cite{mcewen2021machine}.  It was soon realized that the original harmonic mean estimator can fail catastrophically~\cite{neal:1994}, as~it can suffer from an exploding~variance.

The estimator can also be interpreted as importance sampling. Consider the reciprocal marginal likelihood, which may be expressed in terms of the prior and posterior as:
\begin{align}
	\rho
	 & = \int \,\text{d} \theta \:
	\frac{1}{\mathcal{L}(\theta)} \: \prob(\theta | \data)  = \int \,\text{d} \theta \:
	\frac{1}{z} \:
	\frac{\pi(\theta)}{\prob(\theta \given \data)} \:
	\prob(\theta \given \data).
\end{align}
It is clear that the estimator has an importance sampling interpretation where the importance sampling target distribution is the prior $\pi(\theta)$, while the sampling density is the posterior $\prob(\theta | \data)$, in~contrast to typical importance sampling~scenarios.

For importance sampling to be effective, one requires the sampling density to have fatter tails than the target distribution, i.e., to have greater probability mass in the tails of the distribution.  Typically, the prior has fatter tails than the posterior since the posterior updates our initial understanding of the underlying parameters $\theta$ that are encoded in the prior, in~the presence of new data $\data$.
For the harmonic mean estimator, the importance sampling density (the posterior) typically does not have fatter tails than the target (the prior) and so importance sampling is not effective.
This explains why the original harmonic mean estimator can be~problematic.

In~\cite{gelfand1994bayesian}, an arbitrary density $\varphi(\theta)$ is introduced to relate the reciprocal of the marginal likelihood to the likelihood through the following expectation:
\begin{equation}
	\rho
	=
	\mathbb{E}_{\prob(\theta | \data)} \biggl[
		\frac{\varphi(\theta)}{\mathcal{L}(\theta) \pi(\theta)}
		\biggr].
\end{equation}
The above expression motivates the estimator:
\begin{equation}
	\label{eqn:harmonic_mean_retargeted}
	\hat{\rho} =
	\frac{1}{N} \sum_{i=1}^N
	\frac{\varphi(\theta_i)}{\mathcal{L}(\theta_i) \pi(\theta_i)} ,
	\quad
	\theta_i \sim \prob(\theta | \data).
\end{equation}
The normalized density $\varphi(\theta)$ can be interpreted as an alternative importance sampling target distribution, hence we refer to this approach as the re-targeted harmonic mean estimator.  Note that the original harmonic mean estimator is recovered for the target distribution $\varphi(\theta) = \pi(\theta)$.

The learned harmonic mean estimator is introduced in~\cite{mcewen2021machine}, where the target density $\varphi(\theta)$ is learned by machine learning techniques.  It is shown in~\cite{mcewen2021machine} that the optimal target distribution is the posterior.  Since the target must be normalized, the~normalized posterior is clearly not accessible since its normalizing constant is precisely the term of interest.  The~learned harmonic mean approximates the optimal target of the posterior with a learned model that is normalized.  While the approximation need not be highly accurate, it is critical that the probability mass of the learned distribution is contained within the posterior in order to avoid the exploding variance problem.  In~\cite{mcewen2021machine}, a bespoke optimization problem is introduced when training models in order to ensure this property is satisfied.  Specifically, the~model is fitted by minimizing the variance of the resulting estimator, while ensuring it is also unbiased, and~with possible regularization.  Such an approach requires a careful selection of an appropriate model and its hyperparameters for a problem at hand, determined by cross-validation.  Furthermore, only simple classical machine learning models were considered in~\cite{mcewen2021machine}, which in many cases struggle to scale to high-dimensional~settings.

\section{Learning the Target Distribution Using Normalizing~Flows}
\label{sec:method}

In this paper, we learn the target distribution of the learned harmonic mean estimator~\cite{mcewen2021machine} using normalizing flows.  Using normalizing flows renders the previous bespoke approach to training no longer necessary since it provides an elegant way to ensure that the probability mass of the learned distribution is contained within the posterior, thereby resulting in a learned harmonic mean estimator that is more flexible and robust.  Furthermore, normalizing flows also offer the potential to scale to higher dimensional settings.
We first introduce normalizing flows, before~describing how they may be used for the learned harmonic mean estimator and their main advantages in this~context.

\subsection{Normalizing~Flows}
\label{sec:NF_background}

Normalizing flows are a class of probabilistic models that allow one to evaluate the density of and sample from a learned probability distribution (for a review, see~\cite{papamakarios2021normalizing}). They consist of a series of transformations that are applied to a simple base distribution.  A~vector $\theta$ of an unknown distribution $p(\theta)$ can be expressed through a transformation $T$ of a vector $z$ sampled from a base distribution $q(z)$:
\begin{equation}
	\theta = T(z), \text{ where } z \sim q(z).
\end{equation}
Typically, the base distribution is chosen so that its density can be evaluated simply and so that it can be sampled from easily. Often a Gaussian is used for the base distribution.
The unknown distribution can then be recovered by the change of variables formula:
\begin{equation}
	p(\theta) = q(z) \vert \det J_{T}(z)  \vert^{-1},
\end{equation}
where $J_{T}(z)$ is the Jacobian corresponding to transformation $T$. In~a flow-based model, $T$ consists of a series of learned transformations that are each invertible and differentiable, so that the full transformation is also invertible and differentiable.~This allows us to compose multiple simple transformations with learned parameters, into~what is called a flow, obtaining a normalized approximation of the unknown distribution that we can sample from and evaluate.  Careful attention is given to construction of the transformations such that the determinant of the Jacobian can be computed~easily.

A relatively simple example of a normalizing flow is the real-valued non-volume preserving (real NVP) flow introduced in~\cite{dinh2016density}.  It consists of a series of bijective transformations given by affine coupling layers. Consider the $D$ dimensional input $z$, split into elements up to and following $d$, respectively, $z_{1:d}$ and $z_{d+1:D}$, for~$d<D$.  Given input $z$, the~output $y$ of an affine couple layer is calculated by
\begin{align}
	y_{1:d} =   & z_{1:d} ;                                                 \\
	y_{d+1:D} = & z_{d+1:D} \odot \exp\bigl(s(z_{1:d})\bigr) +  t(z_{1:d}),
\end{align}
where $\odot$ denotes Hadamard (elementwise) multiplication.
The scale $s$ and translation $t$ are typically represented by neural networks with learnable parameters that take as input $z_{1:d}$.  This construction is easily invertible and ensures the Jacobian is a lower-triangular matrix, making its determinant efficient to~calculate.

\subsection{Concentrating the Probability Density for the Learned Harmonic Mean~Estimator}
\label{sec:temperature}

Normalizing flows meet the core requirements of the learned target distribution of the learned harmonic mean estimator: namely, they provide a normalized probability distribution for which one can evaluate probability densities. In~this work, we use them to introduce an elegant way to ensure the probability mass of the learned distribution is contained within the posterior. We thereby avoid the exploding variance issue of the original harmonic mean estimator and can evaluate the marginal likelihood accurately without the need for~fine-tuning.

\begin{figure}[H]
	\vspace{-3pt}
	\hspace{-6pt} \includegraphics[width=8cm]{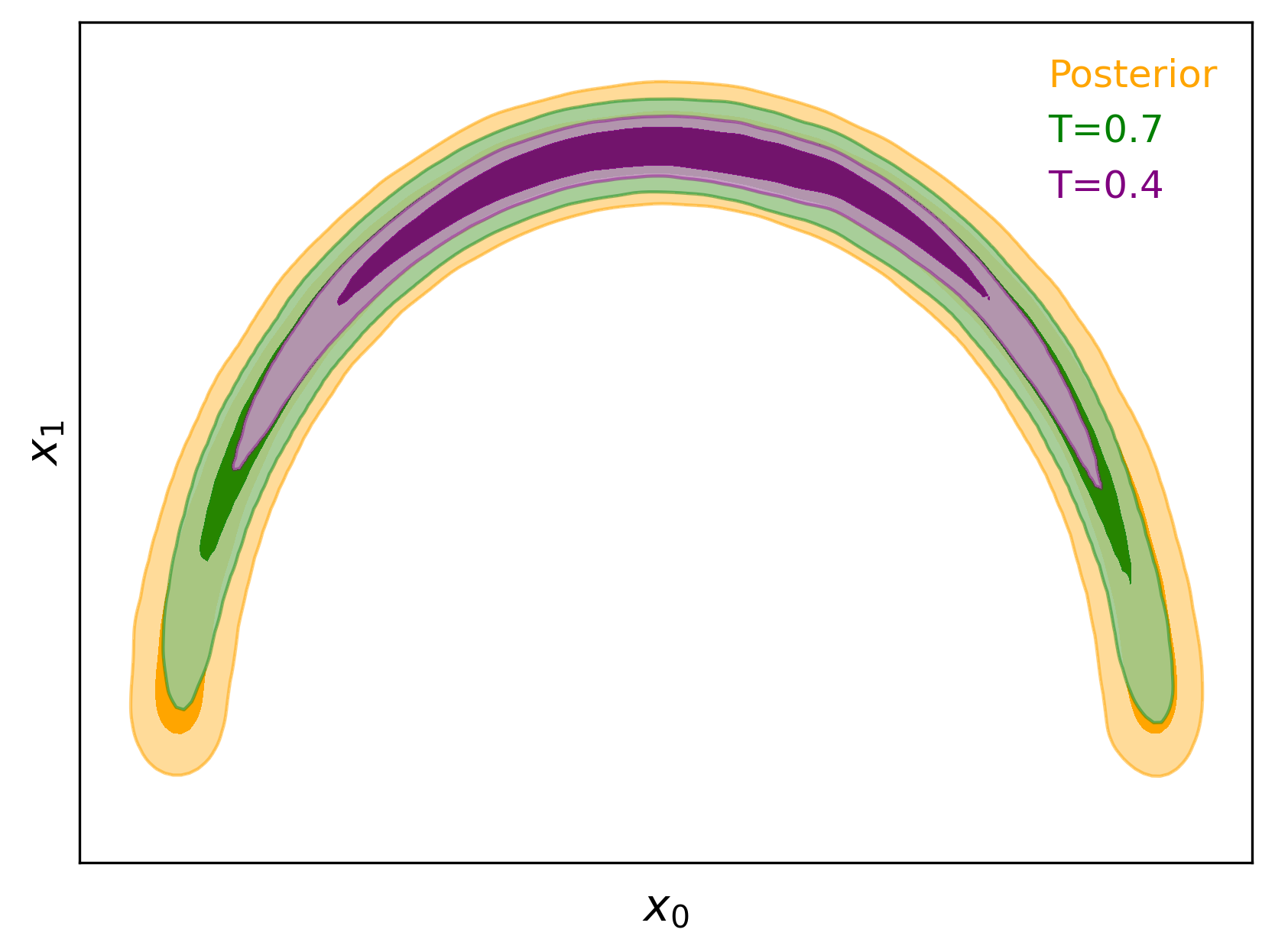}
	\caption{Diagram illustrating the concentration of the probability density of a normalizing flow.  The~flow is trained on samples from the posterior, giving us a normalized approximation of the posterior distribution. The~temperature of the base distribution $T \in (0,1)$ is reduced, which concentrates the probability density of the transformed distribution, ensuring that it is contained within the posterior.  The~concentrated flow can then be used as the target distribution for the learned harmonic mean estimator, avoiding the exploding variance issue of the original harmonic mean~estimator.}
	\label{fig:temperature_diagram}
\end{figure}

Reducing the variance of the base distribution, or~equivalently lowering its ``temperature'' in a statistical mechanics perspective, clearly concentrates the probability density of the base distribution.  This has the effect of also concentrating the probability density of the transformed distribution due to the continuity and differentiability of the flow.  Consequently, once a flow is trained to approximate the posterior, by~lowering the temperature of the base distribution (i.e., \ reducing its variance) we can concentrate the learned distribution to ensure its probability mass is contained within the posterior, as~illustrated in Figure~\ref{fig:temperature_diagram}.

The learned distributions considered previously for the learned harmonic mean estimator~\cite{mcewen2021machine} required the introduction of a bespoke optimization problem for training in order to ensure the learned target is contained within the posterior.  This requires careful selection of an appropriate model and its hyperparameters, determined by cross-validation.  The~introduced normalizing flow approach renders bespoke training no longer necessary.  Instead, we train a flow in the usual manner, based on maximum likelihood estimation, before~concentrating its probability density.  There is only one parameter to consider, the~temperature $T \in (0,1)$.  Moreover, we expect a common value of $T \sim 0.9$ to be suitable for most problems.  Once we have a flow with its probability density concentrated for $\varphi(\theta)$, the~learned harmonic mean estimator can be computed in the usual manner~\cite{mcewen2021machine}.  Using normalizing flows with the learned harmonic mean thus provides a much more robust method.
Furthermore, an~added benefit of using flows is that we can draw samples from the flow distribution efficiently, in~order to easily visualize the concentrated target distribution and compare it to the~posterior.

In this preliminary work, we consider real NVP flows only, as~described above, which are implemented in the harmonic code.  In~the future, we will consider more expressive and scalable flows.  The~use of normalizing flows for the learned distribution therefore has the potential to extend the learned harmonic mean estimator to problems with complex, high-dimensional~posteriors.

\section{Experiments}
To validate the effectiveness of the method described in Section~\ref{sec:method}, we repeat a number of the numerical experiments carried out in~\cite{mcewen2021machine} but using normalizing flows as the target distribution of the learned harmonic mean estimator.
In the experiments that follow, we consider a real NVP flow where the scale and translation networks of the affine coupling layers are given by two-layer dense neural networks with a leaky ReLU in between.  The~scaling layers additionally include a proceeding softplus activation.~We typically consider a flow with six coupling layers, where typically only the first two include scaling, and~permute elements of the vector between coupling layers to ensure the flow transforms all elements.  We consider a Gaussian base distribution with unit variance.
We use the emcee package~\cite{emcee} to generate MCMC samples from the posterior.
We then train the real NVP flow on half of the samples by maximum likelihood and calculate the marginal likelihood using the remaining samples by the learned harmonic mean estimator with the flow concentrated to temperature $T$. In~all experiments, we consider an identical temperature of $T=0.9$, which works well throughout, demonstrating that $T$ does not require fine-tuning.
We consider a relatively simple flow in this preliminary work and a small number of simple experiments.  In~future work, we will consider more expressible and scalable flows, and~further experiments to thoroughly evaluate the robustness and scalability of the~method.

\subsection{Rosenbrock}
\label{sec:rosenbrock}

\begin{figure}[H]
		\vspace{-3pt}
	\hspace{-8pt} \includegraphics[width=9cm]{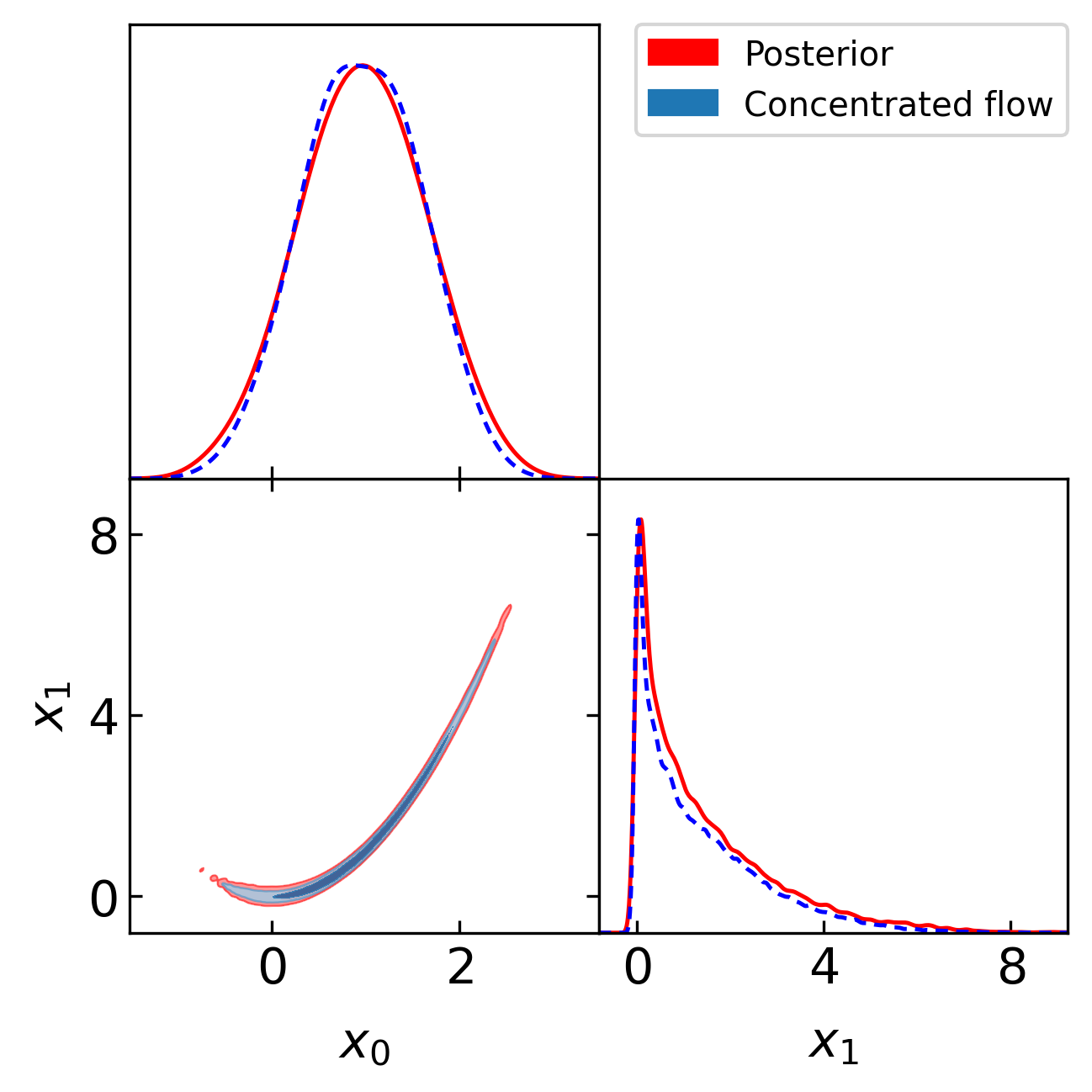}
	\caption{Corner plot of samples from the posterior (red) and real NVP flow with temperature $T=0.9$ (blue) for the Rosenbrock benchmark problem. The~target distribution given by the concentrated flow is contained within the posterior and has thinner tails, as~required for the learned harmonic mean~estimator.}
	\label{fig:rosenbrock_corner}
\end{figure}

A common benchmark problem to test methods that compute the marginal likelihood is a likelihood specified by the Rosenbrock function, which exhibits a narrow curving degeneracy.
We consider the Rosenbrock likelihood in $d=2$ dimensions and a simple uniform prior with $x_0 \in [-10, 10]$ and $x_1 \in [-5, 15]$. We sample the resulting posterior distribution, drawing 5000 samples for 200 chains, with~burn-in of 2000 samples, yielding 3000 posterior samples per chain.
Figure~\ref{fig:rosenbrock_corner} shows a corner plot of the training samples from the posterior (red) and from the normalizing flow (blue) at temperature $T=0.9$.
It can be seen that the concentrated flow approximates the posterior well and has thinner tails, as~required for the marginal likelihood estimate to be stable and~accurate.

\begin{figure}[H]
	\begin{adjustwidth}{-\extralength}{0cm}
		\centering
		\begin{subfigure}{.65\textwidth}
			\centering
			\includegraphics[width=0.8\linewidth]{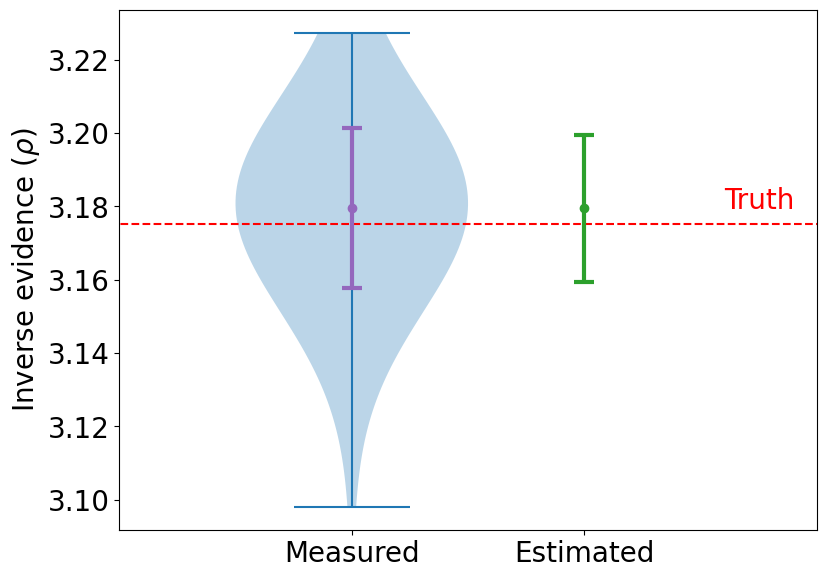}
			\caption{\centering Inverse~evidence}
			\label{fig:rosenbrock_violin_inv_ev}
		\end{subfigure}%
		\begin{subfigure}{.65\textwidth}
			\centering
			\includegraphics[width=0.8\linewidth]{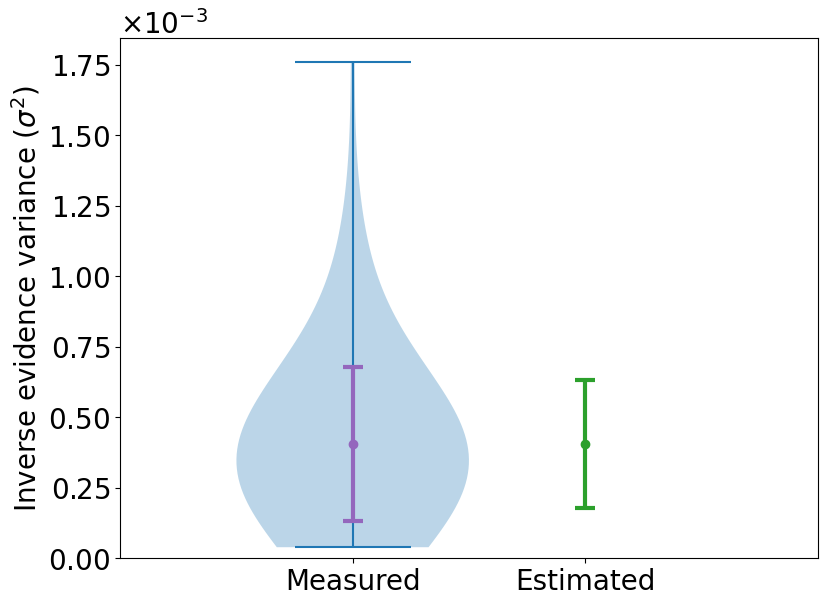}
			\caption{\centering Variance of inverse~evidence}
			\label{fig:rosenbrock_violin_var}
		\end{subfigure}
	\end{adjustwidth}
	\caption{\textls[-5]{Marginal likelihood computed by the learned harmonic mean estimator with a concentrated flow for the Rosenbrock benchmark problem.
		One hundred experiments are repeated to recover empirical estimates of the statistics of the estimator.
		In panel (\textbf{a}), the distribution of marginal likelihood values are shown (measured) along with the estimate of the standard deviation computed by the error estimator (estimated). The~ground truth is indicated by the red dashed line.
		In panel (\textbf{b}), the distribution of the variance estimator is shown (estimated) along with the standard deviation computed by the variance-of-variance estimator (estimated). The~learned harmonic mean estimator and its error estimators are highly accurate.}}
	\label{fig:rosenbrock_violin}
\end{figure}

This process is repeated $100$ times and the marginal likelihood is computed for each trial. Figure~\ref{fig:rosenbrock_violin} shows a summary of the estimates across all the runs. The~dashed red line in
Figure~\ref{fig:rosenbrock_violin}a indicates the ground truth computed through numerical integration, which is tractable in two dimensions.
It can be seen that the learned harmonic mean estimator using a real NVP flow provides an accurate and unbiased estimate of the marginal~likelihood.

\vspace{-2pt}
\unskip

\vspace{-2pt}

\unskip

\subsection{Normal-Gamma}

We consider the Normal-Gamma example, for which the marginal likelihood can be computed analytically~\cite{friel2012estimating,mcewen2021machine}.  It was found that the marginal likelihood values computed by the original harmonic mean estimator do not vary with a varying prior~\cite{friel2012estimating}, highlighting this example as a pathological failure of the original harmonic mean estimator.  We consider the same pathological example here and demonstrate that our learned harmonic mean estimator with normalizing flows is highly accurate (as is the learned harmonic mean with other models;~\cite{mcewen2021machine}).
We consider the {Normal-Gamma model} \cite{bernardo1994bayesian,mcewen2021machine} with data $y_i \sim \text{N}(\mu, \tau^{-1})$, for~$i \in \{1, \ldots, n\}$, with~mean $\mu$ and precision (inverse variance) $\tau$.  A~normal prior is assumed for $\mu$ and a Gamma prior for $\tau$:
\begin{align}
	\mu  \sim \text{N}\bigl(\mu_0, (\tau_0 \tau)^{-1}\bigr)  , \;
	\tau \sim \text{Ga}(a_0, b_0 ),
\end{align}
with mean $\mu_0 = 0$, shape $a_0 = 10^{-3}$, and rate $b_0 = 10^{-3}$.  The~precision scale factor $\tau_0$ is varied to observe the impact of changing prior on the computed marginal~likelihood.

\begin{figure}[H]
		\vspace{-3pt}
\hspace{-3mm}	\includegraphics[width=9cm]{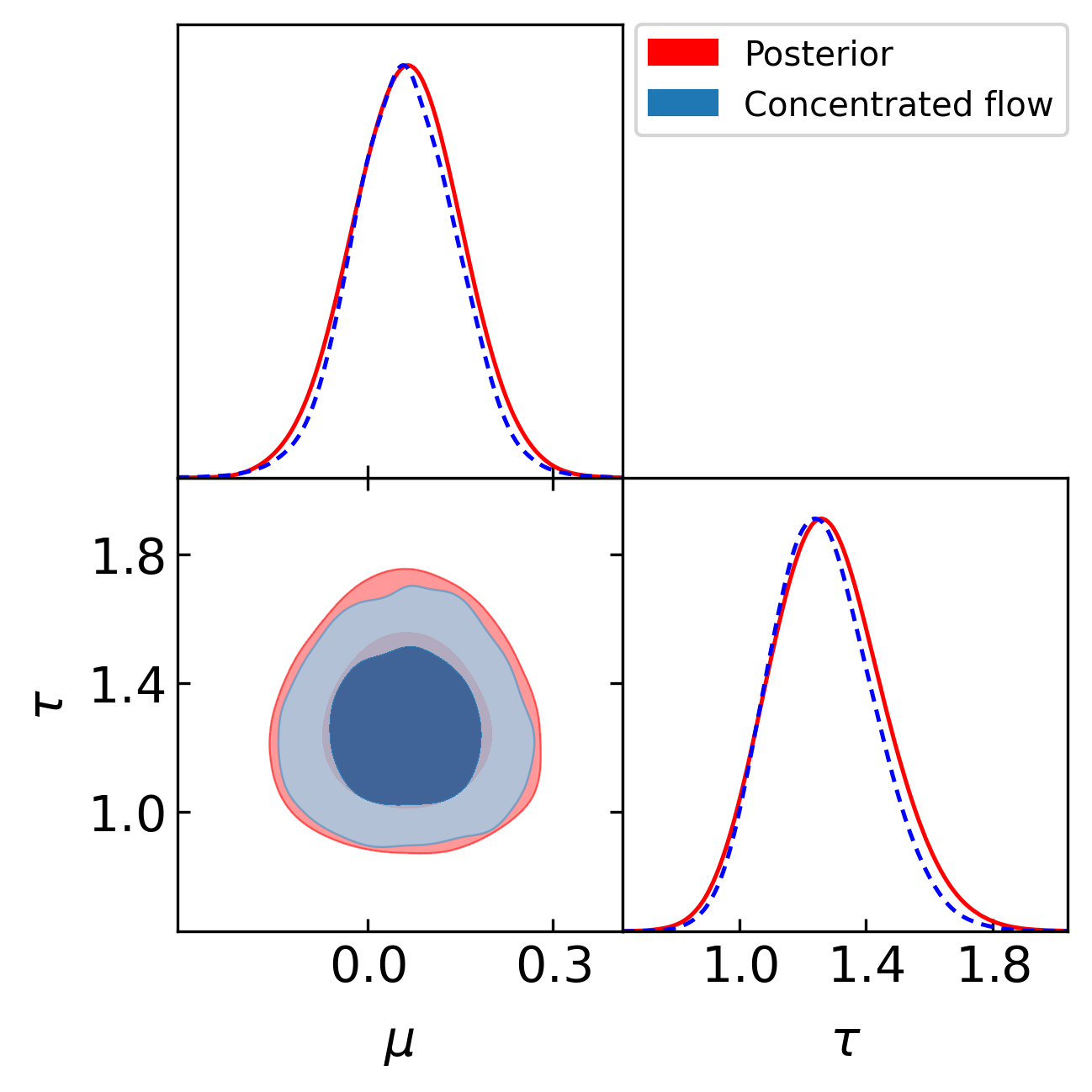}
	\caption{Corner plot of samples from the posterior (red) and real NVP flow trained on the posterior samples with temperature $T=0.9$ (blue) for the Normal-Gamma example with $\tau=0.001$. The~target distribution given by the concentrated flow is contained within the posterior and has thinner tails, as~required for the learned harmonic mean estimator.}
	\label{fig:normalgamma_corner}
\end{figure}

\begin{figure}[H]
	
	\includegraphics[width=9cm]{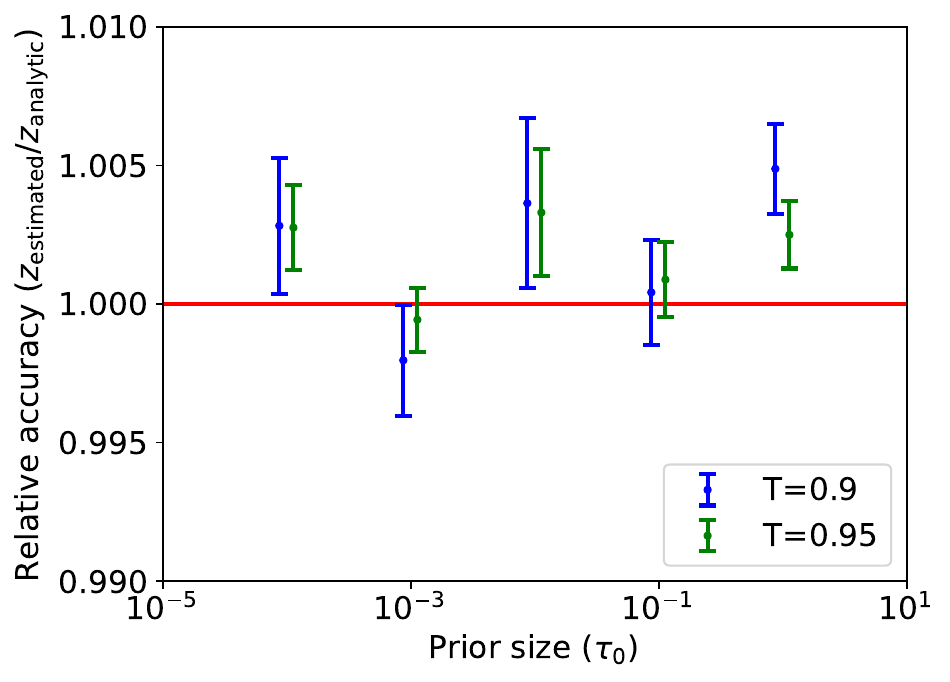}
	\caption{Ratio of marginal likelihood values computed by the learned harmonic mean estimator with a concentrated flow to those computed analytically for the Normal-Gamma problem. Error bars corresponding to the estimated standard deviation of the learned harmonic estimator are also shown. Notice that the marginal likelihood values computed by the learned harmonic mean estimator are highly accurate and are
		indeed sensitive to changes in the prior. Predictions made with flow at temperature $T=0.9$ (blue) and $T=0.95$ (green) are shown, which are slightly offset for ease of visualization, demonstrating that accuracy is not highly sensitive to the choice of $T$.}
	\label{fig:normalgamma_tau}
\end{figure}

We draw 1500 samples for 200 chains, with~burn-in of 500 samples, yielding 1000~posterior
samples per chain.
Figure~\ref{fig:normalgamma_corner} shows a corner plot of the training samples from the posterior for $\tau=0.001$ (red) and from the normalizing flow (blue) at temperature $T=0.9$.
Again, it can be seen the concentrated learned target is close to the posterior but with thinner tails, as~expected.
We consider priors with $\tau \in \left \{ 10^{-4}, 10^{-3}, 10^{-2}, 10^{-1}, 1 \right \}$.
Figure~\ref{fig:normalgamma_tau} shows the relative accuracy of the marginal likelihood computed by the learned harmonic mean estimator using normalizing flows, that is, the ratio of the estimated marginal likelihood to the analytic ground truth.
	We additionally consider a concentrated flow with $T=0.95$ to demonstrate that accuracy is not highly dependent on the temperature parameter.
		It can be seen that the estimate remains accurate and is indeed sensitive to the prior for both temperatures. The~estimates for the flow with $T$ closer to one have a slightly lower variance, as~one would expect, since the broader target $\varphi$ makes more efficient use of~samples.

\vspace{-2pt}

\unskip

\subsection{Logistic Regression Models: Pima Indian~Example}
\label{sec:pima_indian}

We consider the comparison of two logistic regression models using the {Pima Indians} data, which is another common benchmark problem for comparing estimators of the marginal likelihood.  The~original harmonic mean estimator has been shown to fail catastrophically for this example \citep{friel2012estimating}.
The Pima Indians data~\cite{smith1988using}, originally from the National Institute of Diabetes and Digestive and Kidney Diseases, were compiled from a study of indicators of diabetes in $n=532$ Pima Indian women aged 21 or over.  Seven primary predictors of diabetes were recorded, including: number of prior pregnancies (NP);  plasma glucose concentration (PGC); diastolic blood pressure (BP); triceps skin fold thickness (TST); body mass index (BMI); diabetes pedigree function (DP); and age (AGE).
The probability of diabetes $p_i$ for person $i \in \{1, \ldots, n\}$ can be modelled by the logistic function.
An independent multivariate Gaussian prior with precision $\tau=0.01$ is assumed for  parameters $\theta$. Two different logistic
regression models are compared, with~different subsets of covariates:
\begin{align*}
	\text{Model } M_1: & \quad \text{covariates = \{NP, PGC, BMI, DP\} (and bias);}      \\*
	\text{Model } M_2: & \quad \text{covariates = \{NP, PGC, BMI, DP, AGE\} (and bias).}
\end{align*}
A reversible jump algorithm~\cite{green:1995} is used by~\cite{friel2012estimating} to compute a benchmark Bayes factor $\text{BF}_{12}$ of $13.96$ ($\log\text{BF}_{12}=2.6362$), which is treated as ground~truth.

\begin{figure}[H]
	\begin{adjustwidth}{-\extralength}{0cm}
		\centering
		\begin{subfigure}{.67\textwidth}
			\centering
			\includegraphics[width=0.8\linewidth]{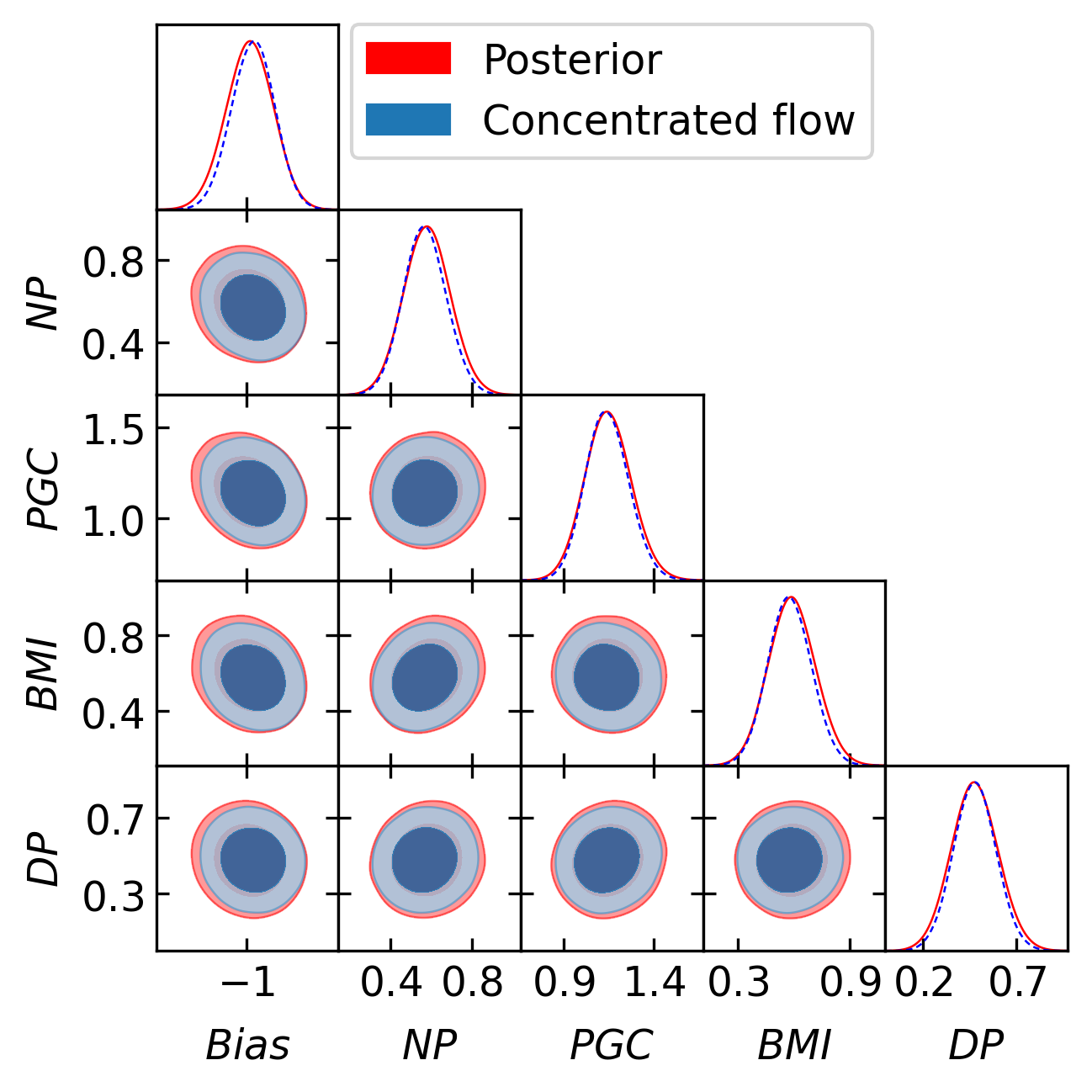}
			\caption{\centering Model~1}
			\label{fig:pima_indian_corner_1}
		\end{subfigure}%
		\begin{subfigure}{.67\textwidth}
			\centering
			\includegraphics[width=0.8\linewidth]{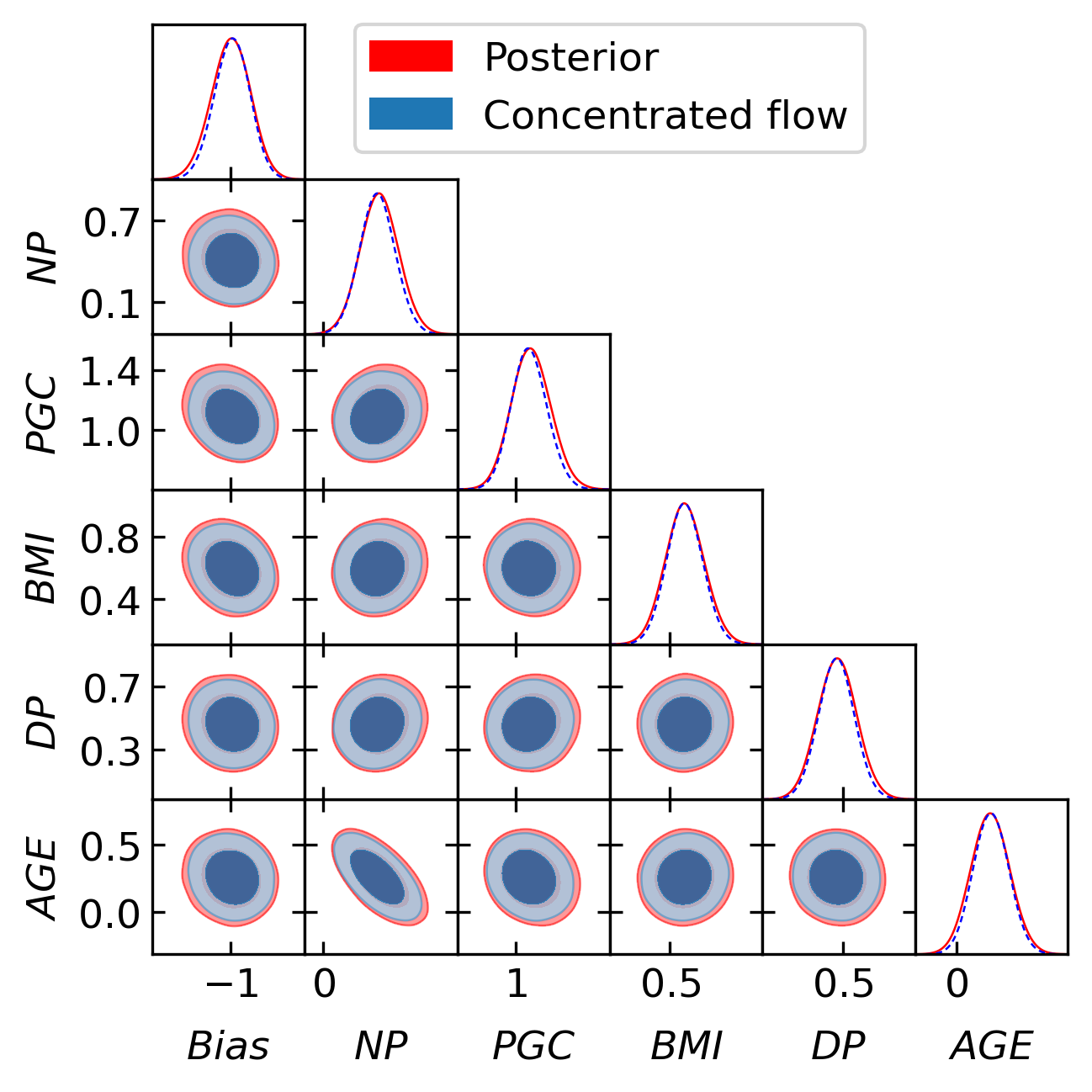}
			\caption{\centering Model~2}
			\label{fig:pima_indian_corner_2}
		\end{subfigure}
	\end{adjustwidth}
	\caption{Corner plot of the samples from the posterior (red) and real NVP flow trained on the posterior samples with temperature $T=0.9$ (blue) for the Pima Indian benchmark problem for $\tau=0.01$.
		The target distribution given by the concentrated flow is contained within the posterior and has thinner tails, as~required for the learned harmonic mean estimator.}
	\label{fig:pima_indian_corner}
\end{figure}

We draw 5000 samples from for 200 chains, with~burn-in of 1000 samples, yielding 4000 posterior
samples per chain. We train a flow consisting of six scaled layers followed by two unscaled ones.
Figure~\ref{fig:pima_indian_corner} shows a corner plot of the training samples from the posterior (red) and from the normalizing flow (blue) at temperature $T=0.9$.
Again, it can be seen that the concentrated learned target is close to the posterior but with thinner tails, as~expected.
We compute the marginal likelihood for Model 1 and Model 2 using our learned harmonic mean estimator.
The log evidence found for Model 1 and 2 is $-257.2300 \pm 0.0020$ and $-259.8602 \pm 0.0031$, respectively, resulting in the estimate $\log\text{BF}_{12}= 2.6302 \pm 0.0051$, which is in close agreement with the~benchmark.

\vspace{-3pt}

\section{Conclusions}

In this work, we propose using normalizing flows for the learned harmonic mean estimator of the marginal likelihood.  The~flow may be fitted to posterior samples by the usual maximum likelihood estimation.  Its probability density may then be concentrated by lowering the temperature of the base distribution, ensuring that the probability mass of the transformed distribution is contained within the posterior to avoid the exploding variance issue of the original harmonic mean estimator.  The~use of flows therefore results in a more robust learned harmonic mean estimator.
We perform a number of experiments to compute the marginal likelihood with the proposed approach, using a real NVP flow, finding excellent agreement with ground truth values.  In~this preliminary work, we consider only simple real NVP flows and a simple set of experiments.  In~a follow-up article, we will consider more expressive and scalable flows to address problems with complex, high-dimensional posteriors.  We will also perform a more extensive set of numerical experiments to thoroughly assess  performance.  This preliminary work nevertheless suggests that the learned harmonic mean estimator with normalizing flows provides an effective technique with which to compute the marginal likelihood for Bayesian model selection.  Furthermore, it is applicable for any MCMC sampling technique or variational inference~approach.

\vspace{6pt}




\authorcontributions{Conceptualization, A.P., M.A.P. and J.D.M.; methodology, A.P., M.A.P. and J.D.M.; software, A.P., M.A.P. and J.D.M.; validation, A.P., M.A.P. and J.D.M.;
	investigation, A.P.; resources, J.D.M.; writing---original
	draft preparation, A.P. and J.D.M.; writing---review and editing, A.P., M.A.P., A.S.M. and J.D.M.; visualization, A.P.; supervision, M.A.P., A.S.M. and J.D.M.;
	funding acquisition, J.D.M. All authors have read and agreed to the published version of the manuscript.}

\funding{A.P.\ is supported by the UCL Centre for Doctoral Training in Data Intensive Science (STFC grant number ST/W00674X/1). A.S.M.\ acknowledges support from the MSSL STFC Consolidated Grant ST/W001136/1. This research was also funded by EPSRC grant number EP/W007673/1.}

\dataavailability{The harmonic code and experiments are available at \url{https://github.com/astro-informatics/harmonic}, accessed on 30 June 2023.}

\conflictsofinterest{The authors declare no conflict of~interest.}

\begin{adjustwidth}{-\extralength}{0cm}

\reftitle{References}

\PublishersNote{}
\end{adjustwidth}

\end{document}